\documentclass[showpacs,preprintnumbers,preprint,aps,prb]{revtex4}
\usepackage[]{amsmath,amssymb,latexsym,epsfig}

\usepackage{graphicx,endnotes,float}
\usepackage{indentfirst}

\begin{document}
\preprint{APS/123-QED}
\title{Structural properties of a calcium aluminosilicate glass from 
molecular-dynamics simulations: A finite size effects study.}

\author{Patrick Ganster}
\affiliation{Laboratoire des Verres UMR 5587, Universit\'e Montpellier II, \\
        Place E. Bataillon, 34095 Montpellier, France \\
        Laboratoire d'\'etude du  Comportement \`a Long Terme, \\
        CEA Valrh\^o, Marcoule, BP 17171, 30207 Bagnols sur C\`eze cedex}

\author{Magali Benoit}
\affiliation{Laboratoire des Verres UMR 5587, Universit\'e Montpellier II, \\
        Place E. Bataillon, 34095 Montpellier, France }

\author{Jean-Marc Delaye}
\affiliation{Laboratoire d'\'etude du  Comportement \`a Long Terme, \\
        CEA Valrh\^o, Marcoule, BP 17171, 30207 Bagnols sur C\`eze cedex}

\author{Walter Kob}
\affiliation{Laboratoire des Verres UMR 5587, Universit\'e Montpellier II, \\
        Place E. Bataillon, 34095 Montpellier, France }

\begin{abstract}
We study a calcium aluminosilicate glass of composition (SiO$_2$)$_{0.67}$-(Al$_2$O$_3$)$_{0.12}$-
(CaO)$_{0.21}$ by means of molecular-dynamics (MD) simulations, using a potential made of two-body 
and three-body interactions. In order to prepare small samples that can subsequently be studied 
by first-principles, the finite size effects on the liquid dynamics and on the glass structural 
properties are investigated. We find that finite size effects affect the Si-O-Si and Si-O-Al angular 
distributions, the first peaks of the Si-O, Al-O and Ca-O pair correlation functions, the Ca 
coordination and the oxygen atoms environment in the smallest system (100 atoms). We give evidence that 
these finite size effects can be directly attributed to the use of three-body interactions.
\end{abstract}

\maketitle

\section{Introduction}

Materials used in nuclear wastes confinement are made of glasses with complex compositions, 
and are based mainly on an alumino-boro-silicate network. These glasses have been studied for 
decades with the aim of testing their durability and their chemical resistance to water 
leaching, the process believed to be the limiting factor for their long time durability 
\cite{aagard,Oelkers,grambow,boksay1,boksay2,Kohn,pacaud,vernaz1,vernaz2}. 
When such glasses are altered by water leaching, the layers formed on the surface are 
hydrated silica-like porous gels enriched with aluminum and calcium. Although it is 
often argued that the protective property of this layer stops the alteration processes induced
by the leaching, no real justification has yet been given \cite{xing,dran,gin1,angeli1, tossel}.

 In calcium aluminosilicate glasses (CAS), silicon and aluminum are both network formers 
and have a tetrahedral coordination with oxygens \cite{wu}. The tetrahedra are connected 
by shared oxygen atoms which are called bridging oxygens (BO) and they
form an amorphous network in which Ca atoms are dispersed. Depending on the Al$_2$O$_3$/CaO 
ratio, Ca atoms play the role of modifiers if they create non-bridging oxygens (NBO) by 
breaking T-O-T linkages (T=Si/Al) and/or they play the role of charge balancing, by neutralizing 
the AlO$_4$ entities.  In principle, 
the number of NBO atoms in a CAS glass  is entirely determined by the Al$_2$O$_3$/CaO ratio 
and this number is directly related to the viscosity of the glass forming liquid. However 
it has been shown that this simple picture is not fully correct and that an excess number
 of NBO atoms exists even for an Al$_2$O$_3$/CaO ratio equal to one, which corresponds - in theory - 
to a perfect polymerized network. This excess
 number of NBO atoms could explain the viscosity anomaly measured in these systems \cite{steb1} 
and could also be related to the existence of other structural units such as oxygen triclusters 
and/or highly coordinated Si and Al atoms. The calcium environment is also 
not very well defined in that it is quite difficult to attribute a specific coordination 
to this species. Nevertheless, in pure CAS systems, it was observed by X-ray and neutron 
spectroscopies that the distribution of oxygen neighbors around Ca atoms is independent 
of the composition \cite{petkov2,cormier}.Another feature observed in aluminosilicate glasses 
is the Al/Al avoidance principle: As the formation of Al-O-Al linkages is energetically less 
favorable than Si-O-Al linkages, the occurrence of the former should be very low in CAS 
structures \cite{meyers, steb2}.

Due to the experimental difficulties to study thin layers, an important first step to obtain
a microscopic understanding of these materials can be achieved by the use of 
molecular-dynamics (MD) simulations. The aim of the present work is therefore to obtain 
a reasonable description of a CAS glass which composition is closed to the composition of 
the protective porous  layer.
However, in the case of such complex ternary systems, it is extremely difficult to find classical
 potentials that are able to give simultaneously a reliable description of the ionic and covalent 
bond lengths, of the bond angles and of the dynamics of the system. Alternatively, one can obtain 
a good description of such structural characteristics and of the dynamics of these complex 
systems from {\sl ab initio} calculations. But in this case, the price to pay is the very limited 
size of the system (100 to 200 atoms) and the very short accessible length of the MD trajectories,
thus making difficult the generation of disordered systems with good statistics on the one hand, 
and the equilibration of the samples at low temperatures on the other hand. 
In the aim of testing the classical potentials, 
we intend to use an approach in which the structure of a glass generated by classical MD is relaxed 
- and therefore tested - by {\sl ab initio} MD. 
This approach has already been successfully used in the case of a SiO$_2$ glass \cite{sio2_epjb} 
and of a Na$_2$O-4SiO$_2$ glass \cite{ns4_jncs}. Since {\sl ab initio} simulations are limited 
to a rather small number of atoms, we first carried out a systematic study of the finite size effects 
on the aluminosilicate system of interest using a classical potential.

In this paper, we present the study of a calcium aluminosilicate glass by MD simulations in 
which we analyzed the dependence of the structural and dynamical properties on the system size. 
In the first section, technical details on the MD simulations are given. Subsequently the 
dynamics of the liquids are presented in section two and the
third section is devoted to the results obtained on the glass structure. 
Finally we discuss and summarize the results.

\section{Methodology}
\label{sec:method}

The chosen composition of the calcium aluminosilicate glass samples for this study is 
67\% SiO$_2$ - 12\% Al$_2$O$_3$ - 21\% CaO in mole percent. 

To describe the interatomic interactions, we used Born-Mayer-Huggins (BMH) and Stillinger-Weber (SW) 
\cite{cheeseman,soule,stillinger} potentials with the modified parameters of Delaye \cite{delaye} which 
were obtained from an earlier study of complex multi-oxide glasses. The
potentials include two-body and three-body terms. In Eq. (\protect\ref{pot2b}) the two-body interaction 
is given, where $r_{ij}$ denotes the distance between atoms $i$ and $j$, and $A_{ij}$, $B_{ij}$, 
$C_{ij}$ and $D_{ij}$ are adjustable parameters. The first term of this pair potential corresponds 
to the Coulomb term, the second one to the core repulsion, and the third and fourth ones are 
referred to as multipolar terms:
\begin{equation}
V_{ij}(r_{ij})= \frac{q_i q_j}{r_{ij}} + A_{ij} \exp( - B_{ij} \; r_{ij} ) 
		- \left( \frac{C_{ij}} {r_{ij}}\right)^6 
		+ \left( \frac{D_{ij}}{r_{ij}}\right)^8.
\label{pot2b}
\end{equation}
The Coulomb interactions were calculated using formal charges $q_i$ equal to +4 for Si, -2 
for O, +3 for Al and +2 for Ca, and the standard Ewald summation method with alpha L=7.2 was used
where $L$ is the size of the cubic box. 
Bond orientations in the CAS system can be described by adding a three-body term involving the 
 O-Si-O, O-Al-O and Si-O-Si triplets. The three-body interaction is given by 
the product of a two-body interaction for each pair of the considered triplets by an angular strained force, 
\begin{equation}
V_{jik}(r_{ij},r_{ik},\theta_{jik}) =
 	\lambda_{jik} \exp{\left( \frac{\gamma_{ij}} {r_{ij} - r_{ij}^0 } +
 	\frac{\gamma_{ik}} {r_{ik} - r_{ik}^0 } \right)} \times
	\left(\cos (\theta_{jik}) -\cos (\theta_{jik}^0)\right)^2
\end{equation}
where $r_{ij}$ and $r_{ik}$ denote the distances between atoms $i$ and $j$ 
(or atoms $i$ and $k$), $\theta_{jik}$ is
the angle between atoms $j$, $i$ and $k$, and $\lambda_{jik}$, $\gamma_{ij}$, $\gamma_{ik}$, $r_{ij}^0$ and
$r_{ik}^0$ are adjustable parameters. 
Tables \ref{tab_pot1} and \ref{tab_pot2} summarize the
parameters of the two-body and the three-body
interactions.
 

 Newton's equations were integrated using a Velocity
Verlet scheme with a time step of 1.8 fs accepting a final deviation of +7 K on the total energy 
for MD runs of 1 million steps. 

Liquid samples of 100 and 200 atoms, generated using a random starting configuration,
 were equilibrated at different temperatures (from $\approx$ 3000 K to $\approx$ 7000 K). 
The lowest temperature for which the silicon self-diffusion was at least 7 \AA\ after 1.8 ns 
of MD trajectory for both system sizes, was taken as the liquid equilibration temperature: 
it corresponds to roughly 4200 K.
Then, we  generated and equilibrated independent liquid samples of different sizes at this temperature 
during 1.8 ns: 8 samples of 100 atoms, 4 samples of 200 atoms, 2 samples of 400 atoms, 2 samples of 800
atoms and 1 sample of 1600 atoms. The sizes of the corresponding boxes were respectively
11.32 \AA, 14.26 \AA, 17.97 \AA, 22.64 \AA\ and 28.52 \AA, corresponding to a density of 
2.42 g.cm$^{-3}$, which is the equilibrium density obtained with this potential at this temperature.
The experimental density for glasses of similar compositions at 300 K lies between 2.55 and
2.60 g/cm$^{3}$ \cite{Doremus,Huang_jncs91}.

\section{Dynamics of the liquid}

\begin{figure}
\begin{center}

\includegraphics[width=18cm]{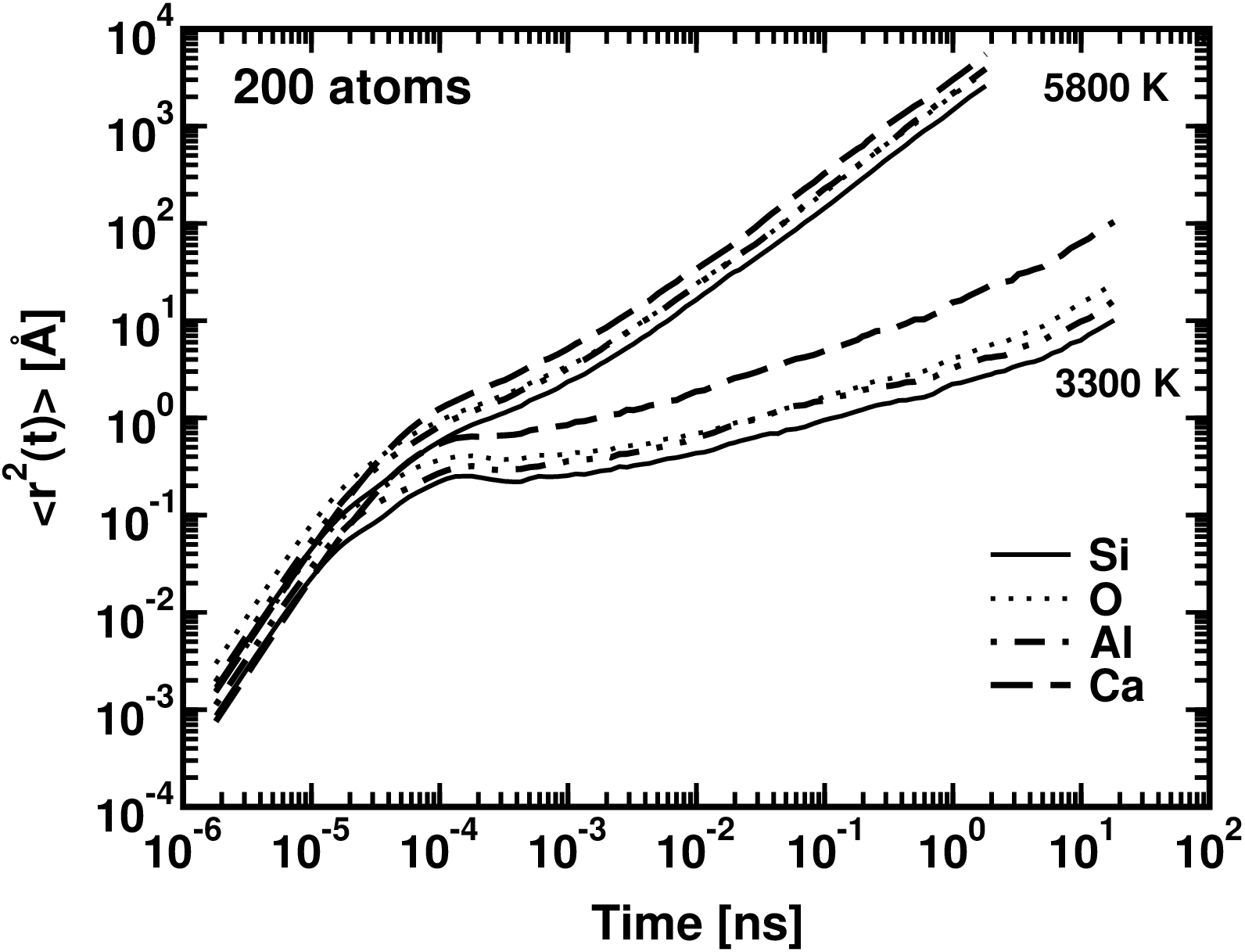}
\caption{ }
\label{figure1}
\end{center}
\end{figure}

The mean square displacements (MSD), $\langle r^2(t) \rangle = \langle \left( r_i(0) 
- r_i(t) \right)^2 \rangle$, of each atomic species were calculated at different 
temperatures  for the systems of 100 and 200 atoms. 
In order to obtain a good description of the dynamical properties,
 it appeared that an average over a larger number of samples was needed. 
Therefore these dynamical properties were studied only for the smallest system sizes 
of 100 and 200 atoms and using averages over 30 independent samples.
 Figure \ref{figure1} represents the  MSD of the various types of atoms 
in the 200-atoms liquid as a function of time in a logarithmic plot. 

At short times, all curves exhibit a straight line with a slope of 2 which indicates 
that all atoms move ballistically. Then, after a transient regime, the atoms show a 
diffusive behavior which is characterized by straight lines with slopes of 1. 
From the mean squared displacements at long times, the self-diffusion constants 
$D$ were extracted using Einstein's law: 
\begin{equation}
  D = \lim_{t\rightarrow \infty }\frac{\langle r^2(t)\rangle}{6t} \ .
\end{equation}
The resulting $D$ of the different species for systems of 200 atoms as a function of 
the inverse temperature are depicted in Fig. \protect\ref{figure2}.
The most diffusive element is Ca, followed by O,  Al and Si. 
It is interesting to note that the Al atoms follow the oxygen self-diffusion for 
all temperatures, despite the fact that, as Si, it is a network former. This behavior resembles the one
found in the Al$_2$O$_3$-SiO$_2$ system \cite{winkler}. Only at the lowest temperature 
the Al self-diffusion approaches the one of the silicon atoms. 
The activation energies of the self-diffusion of the Si, O, Al and Ca atoms were 
extracted from a linear fit of the curves in Fig. \ref{figure2} (for 1/T 
greater than 2.3 $10^{-4}$ K$^{-1}$)
and are equal to 6.4 eV, 5.9 eV, 6.5 eV and 5.3 eV, respectively. 
 
The evolution of the diffusion constants as a function of temperature as well as 
the values of the activation energies do not show any differences between the 
100-atoms and 200-atoms liquids, suggesting that no size effects exist 
- at least up to 200 atoms - for the dynamical quantities.  This is in apparent contrast
with the results of previous studies carried out on supercooled 
liquids \cite{heuer,kim} and in silica \cite{horbach96} 
which shows strong finite size effects in the dynamics of the smallest samples. 
However, in order to study more deeply the finite size effects on the dynamics of the CAS liquids, 
more statistics would  have been needed for systems of larger sizes as well. Since our main purpose 
is to study the size effects on the glass structural properties, we considered that 
this is beyond the scope of the present work. \\

\section{Structural analysis}

After cooling down all liquids from 4200 K to 300 K using a linear quench 
with a rate of $10^{13}$ K.s $^{-1}$ and after a MD simulation of 150 ps at 300 K, 
we performed 1000 additional steps for the structural analysis. \\

\subsection{Radial distribution functions}
\label{sec:RDF}
 
The radial distribution functions (RDF) were evaluated by means of Eq.(\protect\ref{eqgdr}) 
where $N_{\alpha}$ and  $N_{\beta}$ denote the number of atoms for species $\alpha$ 
and $\beta$ and $V$ is the volume \cite{Hansen}:
\begin{equation}
g_{\alpha \beta}(r) = \frac{V}{4 \pi r^2 N_{\alpha}N_{\beta}} \sum_{i \in \{\alpha\}}
\sum_{j \in \{\beta\}} \langle  \delta (r-r_{ij})\rangle
\label{eqgdr}
\end{equation}
Figure \ref{figure3} shows  the radial distribution functions for Si-O, Al-O and Ca-O 
pairs for all system sizes, with the 100-atoms system as dashed-lines. In all systems 
apart from the 100-atoms ones, the positions of the maxima of the Si-O and Al-O RDFs, which 
correspond to the most probable bond distances,  are respectively equal to 1.60 \AA\ 
and 1.76 \AA .  These results are in good agreement with the X-ray diffraction measurements 
by Petkov {\it et al.} \cite{petkov2,petkov1} on CAS glasses of close compositions which 
give values of 1.63 \AA\ and 1.60 \AA\ for Si-O and 1.77 \AA\ and 1.75 \AA\ for Al-O.  
Because of the complexity of the Ca environment, more statistics would have been needed
in order to refine the location of the maximum of the Ca-O RDF first peak. 
The latter can nevertheless be assigned approximately to  2.50 \AA. 
The lower panels in Fig. \protect\ref{figure3} present zooms on the Si-O (a), Al-O (b),
 and Ca-O (c) RDF first peaks. In these zooms, one can clearly notice that the systems of 
100 atoms exhibit a different RDF than the ones of larger sizes since the Si-O and Al-O 
first peaks are narrower and higher.  This indicates that the local order is more 
pronounced in the 100-atoms samples.  The zoom 
on the Ca-O RDF clearly shows  two peaks in the maximum. This double peak is not 
due to poor statistics  but appears to be due to correlations 
that exist in all samples of 100 atoms. 
 
In Fig. \protect\ref{figure4}, we compare the Si-Ca, Al-Ca, Al-Al and Ca-Ca RDFs  for 
the systems of 100- (dashed lines) and 1600-atoms (bold lines). Due to the small number 
of Al and Ca atoms, the RDF curves are quite noisy for both system sizes. The determination 
of interatomic distances is consequently quite difficult,  particularly for the Ca-Ca distance. 
For systems of 100-atoms, we can remark that the RDFs show several peaks that are absent in the 
larger system. Nevertheless, the RDFs of the 100-atoms system follow the global shape of the RDFs of 
the system with 1600 atoms, Fig. \ref{figure4}b. The same behavior is observed for the Si-Si,
 Si-Al and O-O RDFs \cite{ganster_thesis}.

The first peak of the Si-Ca and Al-Ca RDFs of the 1600-atoms system presents a double shoulder, 
Fig. \protect\ref{figure4}a and Fig. \protect\ref{figure4}c, which corresponds to 
two kinds of Si-Ca distances at 3.37 \AA\ and 3.57 \AA. A more pronounced double shoulder 
is also present on the first peak of the Al-Ca RDFs at 3.37 \AA\ and 3.57 \AA\ . 
These double shoulders are due to the presence of non-bridging atoms (NBO) in the SiO$_4$
 or AlO$_4$ groups. As mentioned in the introduction,
the non-bridging oxygen (NBO) atoms  are defined as oxygen atoms that have only one network 
former (Si or Al) as nearest neighbor. Here, these neighbors are counted using cutoff radii 
defined by the location of the minima after the first peaks in the Si-O and Al-O RDFs.

A detailed analysis of the Si-Ca RDF is presented in Fig. 
\protect\ref{figure5}a, b and c.  
 Figure \protect\ref{figure5}a shows the first peak of the Si-O RDF for the different environments
 Q$_x$ of the Si atoms. Since a Q$_x$ atom is defined as a network former atom 
surrounded by $x$ bridging oxygen atoms, a Q$_x$ tetra-coordinated network former 
is thus surrounded by (4-$x$) non-bridging oxygen atoms. Figure  \protect\ref{figure5}a clearly shows 
that the first peak of the Si-O RDF shifts to lower distances when the number of NBO 
is increased on the SiO$_4$ groups (i.e. $x$ is decreased). The Si-O bond length shifts 
from 1.60 \AA\ to 1.52 \AA .  Figure \protect\ref{figure5}b presents the Ca-O RDFs by 
differentiating BO and NBO atoms. A shorter Ca-O bond length is observed when the oxygen 
is non-bridging. By differentiating Q$_x$ Si atoms in the Si-Ca RDFs, Fig. \protect\ref{figure5}c, 
it can be clearly concluded that the first shoulder, observed in the first peak of 
the Si-Ca RDF, is due to NBO atoms and the second to BO atoms, since the separation of the two peaks 
becomes more and more pronounced as the Si atoms are more and more surrounded by NBO atoms. 
          
 From the analysis of the 1600-atoms system, we summarize all interatomic distances of the CAS 
structure in Tab. \ref{tab_dist}, the error bars being of the order of 5$ \cdot 10^{-3}$ \AA\ in the case of 
Si-O and Al-O and between  10$^{-2}$ \AA\ and 5$ \cdot 10^{-2}$ \AA\ for the other distances. 
In the case of Si-Ca and Al-Ca, we took the positions of the 
first peak as a crude estimation of the distances, the RDF first peak being too noisy to allow 
a clear determination of the position of the maximum.
Some of the values given in Tab. \protect\ref{tab_dist} can be compared with experimental 
distances obtained either on glasses or on crystalline analogous. Interatomic distances 
do not change dramatically with  composition and one can compare directly the Si-O and the Al-O  distances 
with the experimental values presented in Tab. \protect\ref{tab_dist}.

In summary, we find clear finite size effects for the 100-atoms system which affect
the first peaks of the Si-O, Al-O and Ca-O RDFs.  These finite size effects suggest that
the glass models become more locally organized  for the smallest system size. 
For  all the other system sizes, the RDFs give information on the local structure
which is overall in agreement with experimental data except for the Ca-O bond length
that is found to be slightly too large.

\subsection{Angular distributions and linkages} 

In order to calculate angular distributions, we  have defined cutoff radii using the location
of the minima after the first peaks in the Si-O and Al-O RDFs, Fig. \protect\ref{figure3}a. 
Figure \ref{figure6} represents the Al-O-Al  (a),  Si-O-Si (b),   Si-O-Al (c),  O-Al-O (d) 
and O-Si-O  (e) angular distributions for the systems of 100 and 1600 atoms. All other 
system sizes show similar distributions as the ones of the 1600-atoms system. Like the RDFs, 
the angular distributions of the systems of 100 atoms follow the general shape of the 1600-atoms 
ones but they present many small peaks that are due to the more ordered character of 
the systems. Particularly in the case of the Si-O-Si angular distribution, we have 
checked that these peaks are not due to statistics but correspond indeed to 
structural correlations  that are present at the same position in all samples of 100 atoms.  
 
In the 1600-atoms case, the O-Al-O and O-Si-O angular distributions, Fig. 
\protect\ref{figure6}d and e, show a quite symmetric shape centered at 
107$\rm^o$ and 108$\rm^o$, respectively, which is in agreement with the values 
measured in amorphous silicates \cite{pettifer}. The Si-O-Si and Si-O-Al angular 
distributions, Fig. \protect\ref{figure6}b and c, have maxima at 157$\rm^o$ 
and 148$\rm^o$ respectively, and exhibit an asymmetric shape with a tail toward 
small angles. The Si-O-Si angles are larger than the Si-O-Al ones which is consistent 
with the results of static {\sl ab initio} calculations by Xiao and Lasaga  \cite{xiao}
in which they showed that the equilibrium Si-O-Si angle of a H$_6$Si$_2$O cluster is larger than 
the Si-O-Al angle of a H$_6$AlSiO cluster.  The Al-O-Al angular distribution, Fig. 
\protect\ref{figure6}a, spreads from 90 to 180$\rm^o$ and does not present
 a distinctive shape with a clear maximum like in the Si-O-Si and Si-O-Al distributions. 
Because of the low number of Al-O-Al linkages in the system, a better statistics would be 
needed in order to refine the Al-O-Al distribution. That a few Al-O-Al linkages are
 present in the glass samples is consistent with the fact that the Al/Al avoidance principle 
\cite{meyers,steb2} is not necessarily fulfilled in amorphous aluminosilicates.  

In order to estimate to what extent the Al/Al avoidance principle is violated in 
our glass samples, we computed the number of T-O-T' linkage concentrations 
(T, T' = Si or Al) that  would be obtained if the Si and Al atoms were randomly 
distributed among all the possible network former sites in the 1600-atoms sample \cite{delaye}.
 We then compared these concentrations with those found in the simulated glasses. 
If the Si$^{+4}$ and Al$^{+3}$ cations were randomly distributed among the network former sites of the CAS glass, 
the number of Al-O-Al linkages would be given by : 
\begin{equation}
N_{\rm AlAl} = N_{\rm O} \frac{N_{\rm AlO} \times (N_{\rm AlO}-1)}{(N_{\rm SiO} + N_{\rm AlO}) \times (N_{\rm SiO} + N_{\rm AlO} -1)}
\label{eqalal}
\end{equation}
where $N_{\rm AlAl}$ is the number of Al-O-Al linkages, $N_{\rm O}$ is the number of 
bridging oxygens, $N_{\rm SiO}$ and $N_{\rm AlO}$ are the number of Si-O and Al-O bonds, respectively.
The probability to find an Al-O-Al linkage corresponds to the probability to pick out 
two Al-O bonds among the different possibilities for the Si-O and Al-O bonds. 
In the system of 1600 atoms, the expectation value for the number 
of Al-O-Al linkages obtained from Eq. (\protect\ref{eqalal}) is 62 whereas only 39 
Al-O-Al linkages are found in the simulated sample. This difference indicates that 
the Al/Al avoidance exists for this kind of glass but is not  completely fulfilled. 

Fig. \ref{figure7} shows the mean number of Si-O-Si, Si-O-Al, Al-O-Al linkages 
as well as the total number of linkages, divided by the total 
number of oxygen atoms, as a function of  the system size. The variation of the number 
of linkages does not exceed 1 \% and remains inside the error bars, thus indicating 
that within the statistical errors no finite size effects can be detected
on this quantity and that size effects only appear on the associated angular distributions.

\subsection{Coordination numbers}

In order to determine the Si, Al and Ca coordination numbers, we used the same cutoff radii 
as the ones used for the angular distributions. We found that the local order of the 
Si and Al atoms is very well defined in that the atoms are always fourfold coordinated whatever 
the size of the system is. However, as the minimum of the Ca-O RDF is not
defined very well, we calculated the Ca coordination distributions for ten different cutoff radii, 
chosen so that they sample regularly the region of the minimum (see inset of 
Fig. \ref{figure8}). Only five of these distributions, as well as their average, 
are shown in Fig. \ref{figure8} for the 1600-atoms system. 
The coordination distribution of the Ca atoms is strongly dependent on the choice of the cutoff
radii and therefore caution should be used while discussing the average Ca coordination number. 

The averaged Ca coordination distributions for the different system sizes are presented 
in Fig. \ref{figure9}. 
From this figure, we can see that the Ca atoms have multiple local environments, 
the number of oxygen surrounding the Ca atoms ranging from 4 to 12. The maxima of these 
distributions are located between 7 and 8 except for the 100-atoms systems. In that case, 
the distribution seems to be shifted toward smaller numbers, to be more asymmetric and 
to have a maximum located at 6. From experimental measurements, the coordination of the 
Ca atoms were estimated to be equal to 7 in WAXS \cite{cormier} and to 5.2 in X-ray 
diffraction \cite{petkov1}, and to 6 by Car-Parrinello simulations in a CAS liquid of 100 atoms 
\cite{benoit}. Therefore it seems that our models slightly overestimate the Ca mean 
coordination and that this overestimation is compensated by finite size effects in the 100-atoms case. 

\subsection{Bridging oxygens, non-bridging oxygens, and oxygen triclusters}

As defined in Sec. \protect\ref{sec:RDF}, the non-bridging oxygen (NBO) atoms  
are oxygen atoms that have only one network former (Si or Al) as nearest neighbor. 
These neighbors are counted using the cutoff radii used 
for the angular distributions as well as for the coordination numbers.

The expected number of NBO atoms in the system can be estimated from simple stoichiometry 
arguments, if the system is supposed to be made of perfect tetrahedra with only 2-fold 
oxygen atoms. In that case, the number of NBO atoms is simply given by:
\begin{equation}
N_{\rm NBO} = 2 \times N_{\rm Ca}- N_{\rm Al}.
\end{equation}
Following this argument, the percentage of NBO atoms in all systems should be equal to 9.52 $\%$, 
if the Ca$^{2+}$ ions fully play their charge balancing role. However, in the simulated samples, 
we always find an excess number of NBO atoms. Fig. \ref{figure10}a shows the percentage of NBO 
as a function of the system size and this percentage is clearly located between 10.7 \% and 
11.8 \%. The excess number of NBO atoms could be due to the use of a too high quench rate, however
recent $^{17}$O NMR experiments on calcium and sodium aluminosilicate glasses \cite{steb1,steblee} 
revealed the presence of excess NBO atoms as well. 

One can observe a slight size dependence for this number indicating that, for larger 
systems, a larger number of NBO atoms is found. It is furthermore interesting to note 
that a large majority of NBO atoms are located on the Si tetrahedra and only a few percent 
of them on Al tetrahedra (between 1.7 and 8.0 $\%$ of the NBO atoms are located on Al tetrahedra). 
This result is in agreement with  X-ray diffraction experiments by Petkov {\it et al.} \cite{petkov1} who 
observed that the NBO atoms are exclusively located on Si tetrahedra in CAS glasses. 
One can also estimate the number of NBO atoms that are located on a given Si tetrahedron, i.e. 
if the Si tetrahedron is a Q$_4$ (no NBO), a $Q_3$ (one NBO), a Q$_2$ (two NBO)  etc. This information 
is given in Tab. \protect\ref{tab:Q_Si}. As it can be seen, there are no clear size effects 
on the relative number of Q$_x$ and most of the Si tetrahedra with NBO atoms  are in the Q$_3$
 conformation. 


It has been proposed that the excess number of oxygen atoms could be compensated by 
oxygen triclusters, i.e. oxygen atoms which are bonded to three network former atoms 
\cite{steb1,benoit}. 
We therefore evaluated the number of oxygen triclusters in our samples and we found that 
the percentage varies between  1 \% and 1.5 \%, Fig. \protect\ref{figure10}c. Again, 
this percentage shows an increase with the system size as for the number of NBO atoms,
Fig. \protect\ref{figure10}a, even if the error bars are quite large for the small 
system sizes. Finally the number of BO atoms (which are not triclusters) was also evaluated 
and the resulting percentage is depicted in Fig. \protect\ref{figure10}b. 

In order to determine if there exists any particular trend in the formation of tricluster 
types (OSi$_3$, OSi$_2$Al, OSiAl$_2$ and OAl$_3$), we evaluated the percentage of these 
units in the case of a random distribution of the Si and Al atoms among the possible
network former sites. 
Let $N_{\rm O3}$ be the number of oxygens triclusters. Then there are 3$\times N_{\rm O3}$ bonds 
between the oxygen triclusters and the three network former neighbors. Among these bonds, 
we can distinguish the number of Si-O3 bonds, given by $N_{\rm Si-O3}$,
and the number of Al-O3 bonds, given by $N_{\rm Al-O3}$.
We consider the entity formed by the oxygen tricluster and its three neighbors as an ensemble 
of three T-O bonds (T = Si or Al). Then, the probability to pick out two Si-O3 bonds 
and one Al-O3 bond among the complete ensemble of 3$\times N_{\rm O3}$ bonds
 can be written for the OSi$_2$Al triclusters, using the standard combinatory notations: 
\begin{equation}
\frac{C^2_{N_{\rm Si-O3}} \times C^1_{N_{\rm Al-O3 }}}{C^3_{3\times N_{\rm O3}} } 
\end{equation}
where 
\begin{equation}
C^n_k = C(n,k)= \frac{n !}{k ! (n-k)!} .
\end{equation}

The derivation of this formula for the other tricluster types is trivial. When we compare 
the expected values for a random distribution of Si and Al among the possible sites to the values obtained
in the simulated sample, an excess of OSiAl$_2$ is observed in the 1600-atoms system since 13 OSiAl$_2$ triclusters 
are found instead of 8.2, the randomly expected. This excess number of OSiAl$_2$ is accompanied 
by a deficit of OSi$_2$Al and OAl$_3$ triclusters, compared to a random distribution of Si and Al. 
This result suggests a possible charge compensation role of the oxygen triclusters and 
could partly explain the violation of the Al/Al avoidance rule since Al-O-Al linkages 
are present in the OSiAl$_2$ triclusters.

\subsection{Rings distribution}

In silicate systems, rings are defined as loops of T-O links (T=Si or Al) and the ring size 
distribution is a measure of the intermediate range order. 
In the case of aluminosilicate glasses, we defined a ring as the smallest loop composed 
of T-O links. Fig. \ref{figure11}  presents the ring size distributions computed 
for all system sizes. The distributions extend from rings of size 3 to rings of size 9 
and they present a maximum located at 6 for all system sizes. As the size of the system 
decreases, the size of the box becomes comparable to the size of the rings and only 
few large rings can be found. However it is interesting to note that the distribution 
maxima do not shift to smaller values for the small systems, which indicates that the 
finite size effects are not significant on this quantity. 
This result is interesting since we know from the RDFs \cite{ganster_thesis} that correlations exist
 in the 1600-atoms system up to 8 \AA\ at least, which is much larger than half the simulation box
length of the 100-atoms system. Yet the fact that the medium range correlations are not taken into
account in the 100-atoms system does not seem to affect the ring size distribution.

As the maximum of the rings distribution is 6 for pure silica \cite{vollmayr} and 5 for aluminosilicate (AS2) 
\cite{winkler}, we do not denote an effect of the presence of aluminum or calcium atoms 
in the rings distribution which is silica-like. 

\subsection{Static structure factor}

In order to compare the global structure of the simulated systems with the one of real systems, 
we computed the neutron static structure factors. This was done  using the relations:
\begin{equation}
S_n(q) = \frac{1}{\sum_{\alpha} N_{\alpha} b_{\alpha}^2} \sum_{\alpha \beta} b_{\alpha}b_{\beta} S_{\alpha
\beta}(q),      
\end{equation}
\begin{equation}
{\rm with} \ \ S_{\alpha \beta}(q) = \frac{f_{\alpha \beta}}{N} \sum_{l=1}^{N_{\alpha}} 
\sum_{m=1}^{N_{\beta}} 
\langle \exp(i{\bf q}\cdot({\bf r}_l - {\bf r}_m)) \rangle 
\end{equation}
where ${\bf q}$ is the scattering vector, $N$ the number of atoms, $N_{\alpha}$ and $N_{\beta}$ 
the number of atoms of species $\alpha$ and $\beta$, respectively.  The factor $f_{\alpha \beta}$ 
is equal to 0.5 for $\alpha\neq\beta$ and equal to 1.0 for $\alpha=\beta$, $b_{\alpha}$ and 
$b_{\beta}$ are the neutron scattering lengths of species $\alpha$ and $\beta$, and are equal 
to 4.149, 5.803, 3.449, 4.700 fm for Si, O, Al and Ca, respectively \cite{Ni95}.

Figure \protect\ref{figure12} shows the comparison of the neutron structure factors calculated 
for the systems of 100 and 1600 atoms. Even if the 100-atoms $S_n$(q) is noisier, there are no
 perceptible differences between the 100- and the 1600-atoms structure factors. From the partial
 structure factors, the principal contributions to the $S_n$(q) first peak appear to be due to 
the Si-Si, Si-O, Al-O and O-O correlations (see Fig. \protect\ref{figure13}). 
Concerning the second peak, the main contributions can be attributed to Si-Si, Si-Al, O-O and Ca-O correlations. It is, however, more difficult to make a clear relation between the other peaks and the partial correlations. 

On the other hand, in the 1600-atoms $S_n(q)$, one can notice the existence of a very small pre-peak around 0.7 \AA$^{-1}$ (denoted by a vertical dashed line in Fig. \protect\ref{figure13}) which appears to be due to Si-Si, Si-O and O-O correlations. Since the Al-Al correlations do not contribute to this pre-peak, it is very likely that this feature is indeed due to the presence of the Ca atoms which induce the 
existence of pockets or channels and consequently, the oorganisation of the matrix on large scales. 
This type of long-range organisation has already been observed in sodo-silicate glasses and melts 
\cite{NS_horbach}.

Since no experimental data were available for the specific composition of the present work, 
in order to compare the computed $S_n$(q) with experimental data, we had to generate a glass 
with a slightly different composition (60\% SiO$_2$, 10\% Al$_2$O$_3$, 30\% CaO). This new 
simulation was performed on a system of 800 atoms following the same procedure as described 
beforehand. The comparison between the computed and experimental structure factors from 
Ref. \cite{cormier} is presented in  Fig. \protect\ref{figure14}. The agreement is good both 
for the positions of the peaks and for their intensities except at large q where irregularities 
 do not permit to superpose the shoulder after the fourth peak.   

The difference between the two computed compositions is the SiO$_2$/CaO ratio, which is lower 
in the case of the experimental comparison. The differences between the structure factors at the two 
compositions, Fig. \protect\ref{figure12} and Fig. \protect\ref{figure14}, mainly concern 
the two first peaks. In the experimental composition the first peak is lower and the second 
one higher than in the initial  studied composition. This shift is due to the larger quantities 
of Ca in the experimental composition and has already been observed in a neutron scattering 
study of calcium aluminosilicate glasses of different compositions \cite{cormier}.  

\section{Discussion}

The set of interatomic potentials used to simulate the calcium aluminosilicate glasses 
leads to structures that reproduce a series of experimental features if the 
studied sample contains at least 200 atoms. 

Firstly, the simulated neutron structure factor is globally in agreement with 
the experimental one although some of the first neighbor distances are not correctly reproduced.
Recent reverse Monte Carlo calculations \cite{delaye} carried out on glasses of several compositions
have suggested that the distance discrepancies in the local ionic environments
 could explain the slight differences between the simulated and experimental structure factors. 
By stretching the local Si-O distances by 0.02 \AA\ and reducing the Al-O and Ca-O distances 
by respectively 0.05 \AA\ and 0.1 \AA, these differences could be  removed. 

Secondly, the Al/Al avoidance rule is qualitatively reproduced but we have no quantitative 
experimental results to compare with simulations. The fact that the avoidance rule is not 
completely obeyed has already been suggested in recent NMR experiments \cite{steblee}. 
The authors predict that there still exist Al-O-Al linkages in a sodium aluminosilicate glass 
with Si/Al = 0.7 and in a calcium aluminosilicate glass with Si/Al = 0.5 which suggests that 
the respective concentrations in Si-O-Si, Si-O-Al and Al-O-Al are dependent on the energetic 
cost of the local configurations surrounding each linkage. 

Finally, another interesting structural feature concerns the existence of oxygen triclusters. 
If we consider that each Ca atom is "used" to compensate the charged AlO$_4$ tetrahedra and that the 
remaining Ca ions create non-bridging oxygens, the predicted concentration of non-bridging 
oxygens is lower than the simulated ones. This result means that the Ca atoms are not completely 
consumed by the AlO$_4$ tetrahedra, and part of the AlO$_4$ tetrahedra are charge-balanced by 
the oxygen triclusters. The existence of the latter, proposed originally in Ref. \cite{lacy}, 
has not been directly observed but is deduced from $^{17}$O NMR measurements in aluminosilicate
 glasses. The existence of 5 \% of non-bridging oxygen in a CaAl$_2$Si$_2$O$_8$ glass, detected
 by $^{17}$O NMR spectroscopy, suggests also the presence of oxygen triclusters \cite{stebga}. 
 Previous simulations based on similar potentials have also predicted the existence of oxygen 
triclusters in a CaAl$_2$Si$_2$O$_8$ glass \cite{nevinspera}.

However for the glass models containing 100 atoms, the structural characteristics are quite
different indicating that these systems are subject to finite size effects. 
To test whether these size effects are due to the used potential or to the complexity of the glass, 
we performed simulations on a 78-atoms system of pure a-SiO$_2$ by using two different potentials: 
the present one detailed in section \protect\ref{sec:method} and the so-called BKS potential 
derived by van Beest {\it al.} \cite{BKS}. 
Whereas the SiO$_2$ glass obtained with the BKS potential did not show any finite size effects, the
size effects observed in the present work on the RDF first peaks and on the angular distributions were
still present with the potential described in section \protect\ref{sec:method}. 
The main differences between the two potentials reside in the fact that
the BKS potential comprises partial charges for the Si and O ions and does not include three body terms.

In order to determine whether the size effects in the CAS system are induced by the presence 
of the three body terms, we have performed calculations without the three body terms. By  neglecting 
the angular terms, we do not obtain a correct structure, since a large number of miscoordinated 
silicon atoms are found. However the structural quantities concerned by the finite size effects 
can still be examined. For instance,  the first peaks of the Si-O 
and Al-O radial distribution functions for systems of 100 and 1600 atoms have been compared
without the three body terms and we clearly observe the disappearance of the size effects on those RDFs.
The finite size effects also disappear in the angular distributions.
These considerations allow us to clarify the effect of the angular terms on the calculated 
glassy structures. The enhancement of the local order observed for the small systems is due 
to the three body terms: it completely disappears when these terms are removed.

However, for the smallest systems, finite size effects are still present since the structure on distances beyond $L/2$ is influenced by the periodic boundary conditions. Therefore it is not possible to investigate the intermediate range order, which in these type of materials extends up to 10-15 \AA , in systems that are smaller than 10 \AA .
These effects are not responsible of the enhancement of the local structure, but rather of the 
modification of the structure factor at low $q$. These effects are somehow trivial and can be 
easily be taken into account.

\section{Conclusion}
  In this work, we have presented a systematic study of the liquid dynamics and the
structural properties of a calcium aluminosilicate glass as a function of the system  size. 
This study was carried out by  means of molecular dynamics simulations using  a potential
 made of two-body and three-body interactions. The results show that finite size effects 
exist for the smallest system size (100 atoms) and affect a large part of structural
 characteristics: the Si-O-Si and Si-O-Al  angular distributions, the Ca coordination numbers, 
the first peaks of the Si-O, Al-O  and Ca-O pair distribution functions, and the relative 
number of bridging oxygens, non-bridging oxygens and oxygens triclusters.
  
Most of these size effects were clearly attributed to the  presence of three-body interactions 
in the potential. Interestingly, no finite size effects were visible in the relative 
percentages of Al-O-Al linkages and on the ring size distribution. For the larger system 
size, the structural characteristics were compared to experimental  data when possible and 
a correct agreement was obtained for the Si-O and Al-O interatomic distances and for the
 neutron static structure factors. However, the Si-O-Si and Si-O-Al  mean angles and 
the Ca coordination numbers were found to be slightly too large. 

In conclusion, the present work gives confidence in the fact that the local structure  
of relatively complex glasses can be studied by the means of {\sl ab initio} simulations 
carried  out on relatively small samples (200 atoms). On the other hand, it also gives 
indications  on the weaknesses of the presently used classical potential and guidelines 
for possible  improvements.

\newpage

\begin{table}
\begin{center}
\caption{\label{tab_pot1} Parameters of the two-body potentials.}
\begin{ruledtabular}
\begin{tabular}{ccccc}
 atomic pair   & $A_{ij}$ [eV]  &  $B_{ij}$ [\AA] & $C_{ij}$ [eV] & $D_{ij}$ [eV] \\
 \hline 
{  Si-Si  }& {875.9321}   &   3.448  &         &           \\
{ Si-O   }& { 1044.9519 }  &   3.049  &         &           \\
{ Si-Al  }& {  956.3850 }  &   3.448  &         &           \\
{ Si-Ca  }& { 4000.4894 }  &   3.448  &         &           \\
{ O-O    }& {  369.2783 }  &   2.857  &         &           \\
{ O-Al   }& { 1725.0881 }  &   3.448  &         &           \\
{ O-Ca   }& { 8973.4045 }  &   3.448  & 539.9580 &  884.2333  \\
{ Al-Al  }& { 1039.5842 }  &   3.448  &         &           \\
{ Al-Ca  }& { 4326.2957 }  &   3.448  &         &           \\
{ Ca-Ca } & { 17897.5012 }  &   3.448  &         &           \\
\end{tabular}
\end{ruledtabular}
\end{center}
\end{table}
\vspace{2cm}
\begin{table}
\begin{center}
\caption{\label{tab_pot2} Parameters of the three-body potentials.}
\begin{ruledtabular}
\begin{tabular}{ccccccc}
 atomic triplet   & $\lambda_{jik}$ [eV] & $\gamma_{ij}$ [\AA] & $\gamma_{ik}$ [\AA] & $r_{ij}^0$ [\AA] & $r_{ij}^0$
[\AA] & $\theta_{jik}$ [degree]\\
\hline
  O-Si-O   &   149.7960   &   2.6  & 2.6    &  3.0 & 3.0   & 109.5    \\
  O-Al-O   &   149.7960   &   2.6  & 2.6    &  3.0 & 3.0   & 109.5    \\
  Si-O-Si  &   6.2415     &   2.0  & 2.0    &  2.6 & 2.6   & 160.0    \\
\end{tabular}
\end{ruledtabular}
\end{center}
\end{table}
\vspace{2cm}
\begin{table}
\begin{center}
\caption{Interatomic distances in the CAS glass for a system of 1600 atoms
at 300 K, as extracted from the positions of the first peak maxima in 
the corresponding $g_{\alpha \beta}(r)$.}
\begin{ruledtabular}
\begin{tabular}{ccc}
                 &   \multicolumn{2}{c}{Interatomic distance [\AA]}  \\        
 atomic pair     &    This work         &  Experimental 		      \\
		 \hline
 Si-Si     	 &     3.18   		&  3.09\footnotemark[1]          \\
 Si-O            &     1.60  	        &  1.63\footnotemark[2], 1.60\footnotemark[3]  \\
 Si-Al     	 &     3.24   		&	       	   	      \\
 Si-Ca     	 &     3.50   		& 				\\
 O-O       	 &     2.59   		&   2.65\footnotemark[1]       \\
O-Al      	 &     1.76   		&   1.77\footnotemark[2], 1.75\footnotemark[3]$^,$\footnotemark[4] , 1.74\footnotemark[5]         \\
 O-Ca      	 &     2.50   		& 2.32\footnotemark[3] , 2.40\footnotemark[6]     \\
 Al-Al     	 &     3.25   		&  				\\
 Al-Ca     	 &     3.50   		& 				\\
 Ca-Ca     	 &       -    		&               		\\
\end{tabular}
\end{ruledtabular}
\label{tab_dist}
\end{center}
\footnotetext[1] {Reference \cite{himmel}}
\footnotetext[2] {Reference \cite{petkov1}}
\footnotetext[3] {Reference \cite{petkov2}}
\footnotetext[4] {References \cite{mcmillan}}
\footnotetext[5] {Reference \cite{calas}}
\footnotetext[6] {Reference \cite{cormier}}
\end{table}

\vspace{2cm}

\begin{table}
\begin{center}
\caption{Percentages of the Si tetrahedra in the Q$_x$ environment for all
system sizes.} 
\begin{ruledtabular}
\begin{tabular}{crrrrr}
 & \multicolumn{5}{c}{System size} \\
 & 100 & 200 & 400 & 800 & 1600 \\
\hline
Q$_1$ &	0       & 0.6   & 0     &  0.6  & 0 \\
Q$_2$ &	2.9	& 1.7   & 2.3   &  3.4 & 3.7 \\
Q$_3$ & 22.7	& 23.8	& 26.1  & 23.9 & 25.3 \\
Q$_4$ & 74.4	& 73.9	& 71.6	& 72.1 & 71.0  \\
\end{tabular}
\end{ruledtabular}
\label{tab:Q_Si}
\end{center}
\end{table}

\clearpage


\centerline{FIGURE CAPTIONS} 

\noindent {\sc Figure 1~:} Time and temperature dependence of the mean squared
displacements of Si, O, Al and Ca atoms for the 200-atoms systems at 3300 K and 5800 K. \\

\noindent {\sc Figure 2~:} Inverse temperature dependence of the self diffusion
constants for Si, O, Al, and Ca atoms in the 200-atoms liquids. \\

\noindent {\sc Figure 3~:} {\bf (a)} Si-O, Al-O and Ca-O radial distribution functions for systems of 100 (dashed lines), 200, 400, 800 and 1600 atoms (bold lines). 
{\bf (b)} Zoom on the first peak of the Si-O radial distribution function.
{\bf (c)} Zoom on the first peak of the Al-O radial distribution function.
{\bf (d)} Zoom on the first peak of the Ca-O radial distribution function. \\

\noindent {\sc Figure 4~:} {\bf (a)} Si-Ca, {\bf (b)} Ca-Ca, {\bf (c)} Al-Ca and {\bf (d)} Al-Al radial distribution functions
for the systems of 100 (dashed lines) and 1600 atoms (bold lines). \\

\noindent {\sc Figure 5~:} {\bf (a)} Si-O radial distribution functions for different Si environments {\sl Q$_2$}, {\sl Q$_3$}, {\sl
Q$_4$}  and for all Si atoms. {\bf (b)} Ca-O radial distribution functions for different oxygen atom types 
BO, NBO and for all oxygen atoms. {\bf (c)} Si-Ca radial distribution functions for different Si environments  {\sl Q$_2$}, {\sl
Q$_3$}, {\sl Q$_4$} and for all Si atoms. All RDFs are computed for the 1600-atoms system. \\

\noindent {\sc Figure 6~:} Angular distributions of  the {\bf (a)} Al-O-Al, {\bf (b)} Si-O-Si, {\bf (c)} Si-O-Al, 
{\bf (d)} O-Al-O and {\bf (e)}  O-Si-O  
angles for  systems of 100 (dashed lines) and 1600 (bold lines) atoms. \\

\noindent {\sc Figure 7~:} Number of linkages divided by the number of oxygen atoms as the function of the
system size: {\bf (a)} Si-O-Si, {\bf (b)} Si-O-Al, {\bf (c)} Al-O-Al and {\bf (d)} sum of Si-O-Si, Si-O-Al and Al-O-Al. \\

\noindent {\sc Figure 8~:} Averaged probability distribution for finding N oxygen atoms around
the calcium atoms in a system of 1600 atoms and for different 
cutoff radii. The inset shows schematically how the cutoff radii were chosen 
around the minimum after the first peak of the Ca-O radial distribution function. \\

\noindent {\sc Figure 9~:} Probability distribution for finding N oxygen atoms
around the calcium atoms for systems from 100 to 1600 atoms.\\

\noindent {\sc Figure 10~:} Concentration of  {\bf (a)} NBO atoms, {\bf (b)} BO atoms and {\bf (c)} oxygen triclusters as the function of the system size. \\

\noindent {\sc Figure 11~:} Probability of finding rings of N tetrahedra for systems of 100, 200, 400, 800 and 1600 atoms. \\

\noindent {\sc Figure 12~:} Static neutron structure factors for 100-atoms (bold line) and 1600-atoms (circles) systems.  The
scattering lengths were taken to be equal to $b_{Si, O, Al, Ca}$ = 4.149, 5.803, 3.449, 4.700  fm. \\

\noindent {\sc Figure 13~:} Partial structure factors for the 1600-atoms system. 
The curves have been shifted upward and downward for clarity and the vertical dashed line denotes the position of the pre-peak. \\

\noindent {\sc Figure 14~:} Comparison of the experimental (from Ref. \protect\cite{cormier}) and calculated neutron structure factors for a 800-atoms CAS glass of composition 60\% SiO$_2$ - 10\% Al$_2$O$_3$ - 30\% CaO. The scattering lengths were taken to be equal to $b_{Si, O, Al, Ca}$ = 4.149, 5.803, 3.449, 4.700  fm. \\

\newpage

\begin{figure}
\begin{center}
\includegraphics[width=18cm]{figure2.eps}
\caption{ }
\label{figure2}
\end{center}
\end{figure}

\newpage

\begin{figure}
\begin{center}
\includegraphics[width=18cm]{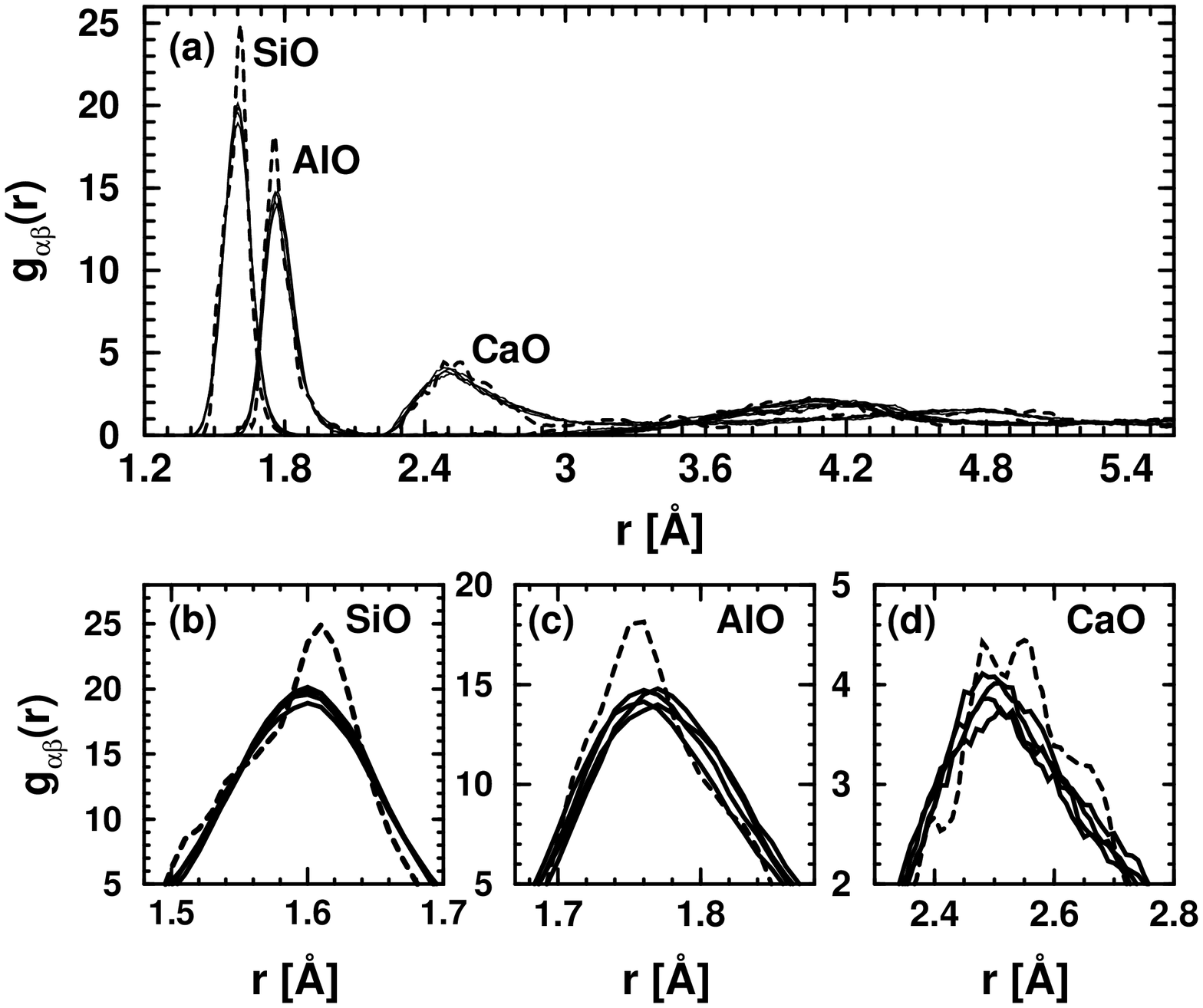}
\caption{ }
 \label{figure3}
\end{center}
\end{figure} 

\newpage

\begin{figure} 
\begin{center}
\includegraphics[width=18cm]{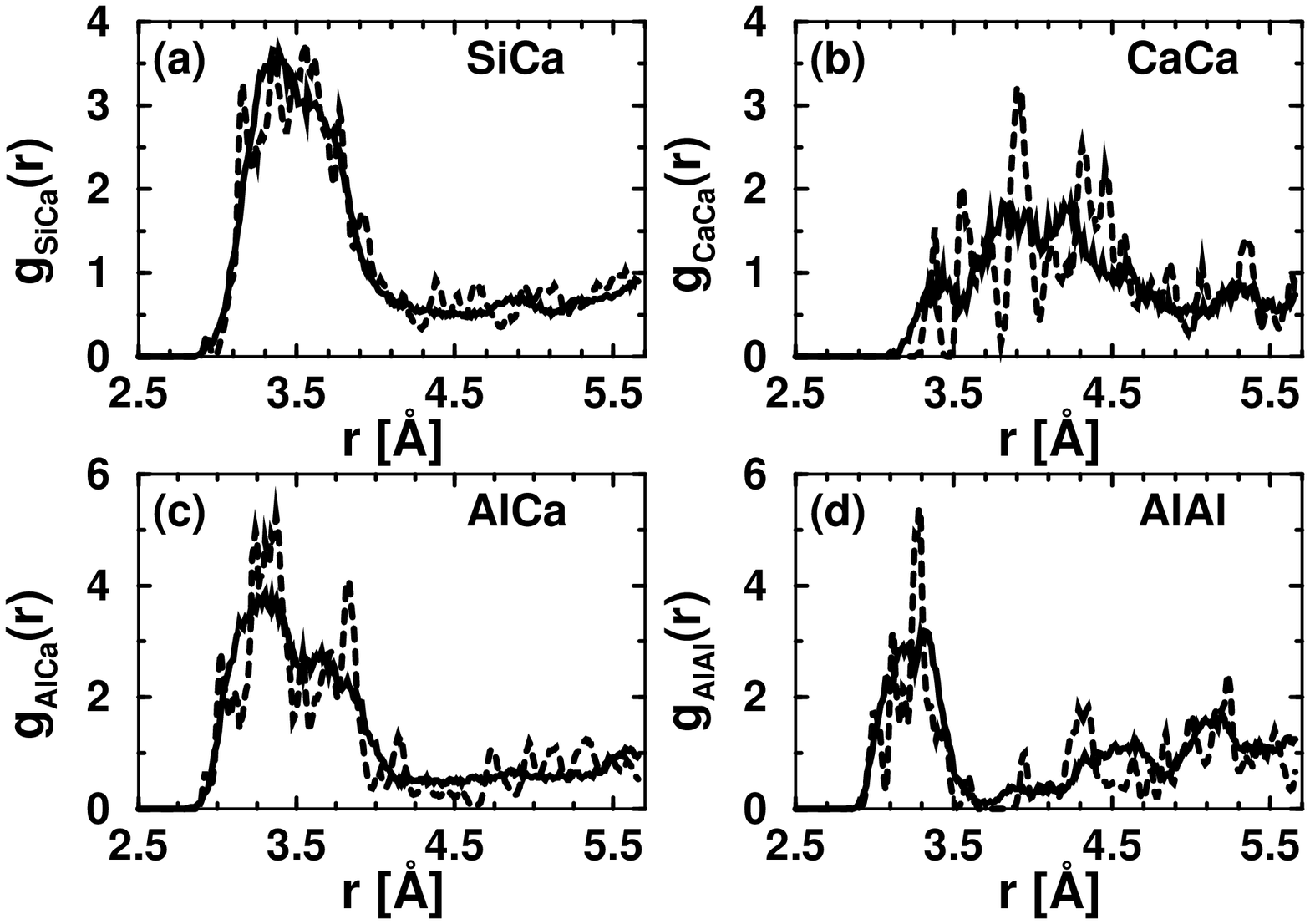}
\caption{ } 
\label{figure4}
\end{center}
\end{figure}

\newpage

\begin{figure}
\begin{center}
\includegraphics[width=14cm]{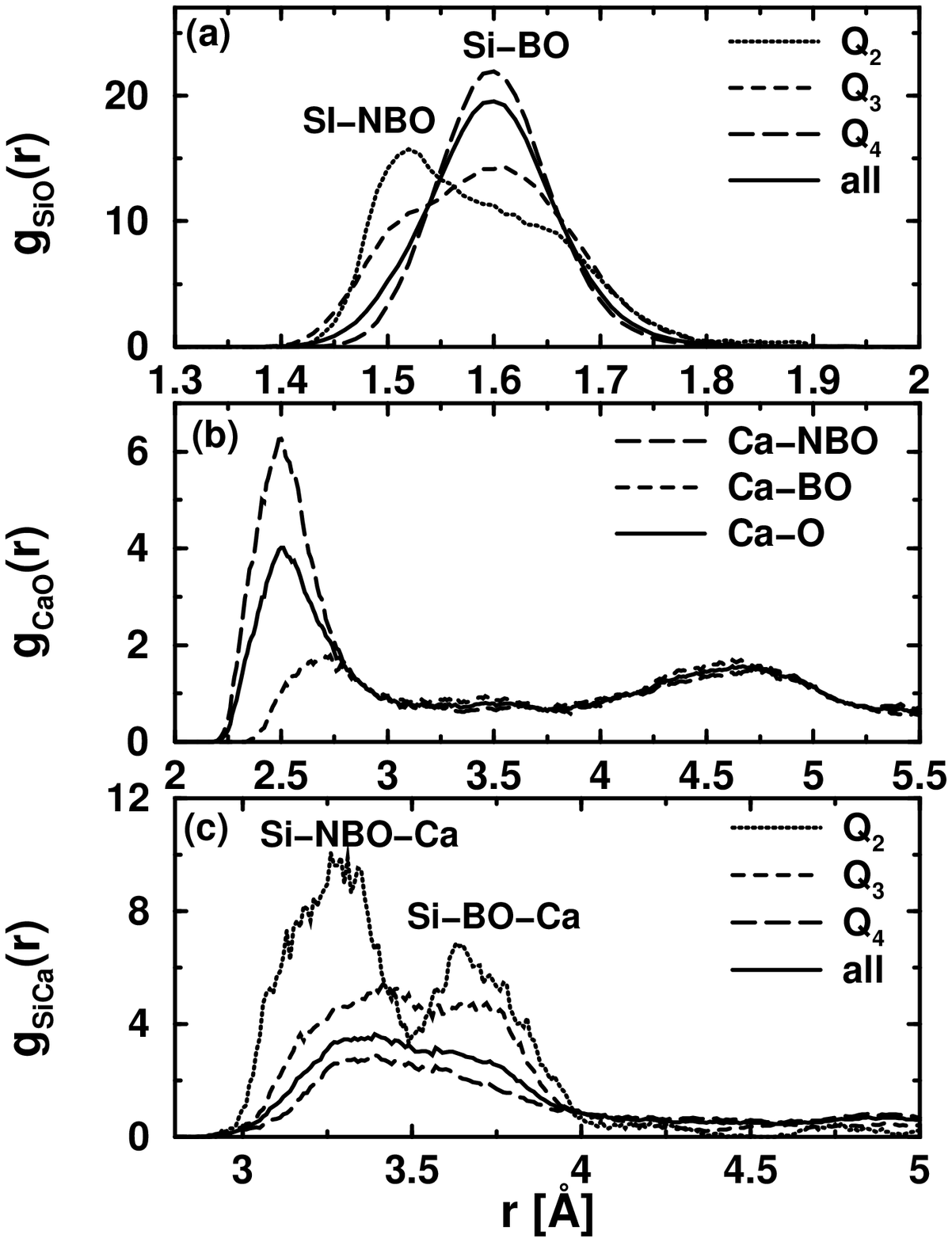}
\caption{ }
\label{figure5}
\end{center}
\end{figure}

\newpage

\begin{figure}
\begin{center}
\includegraphics[width=18cm]{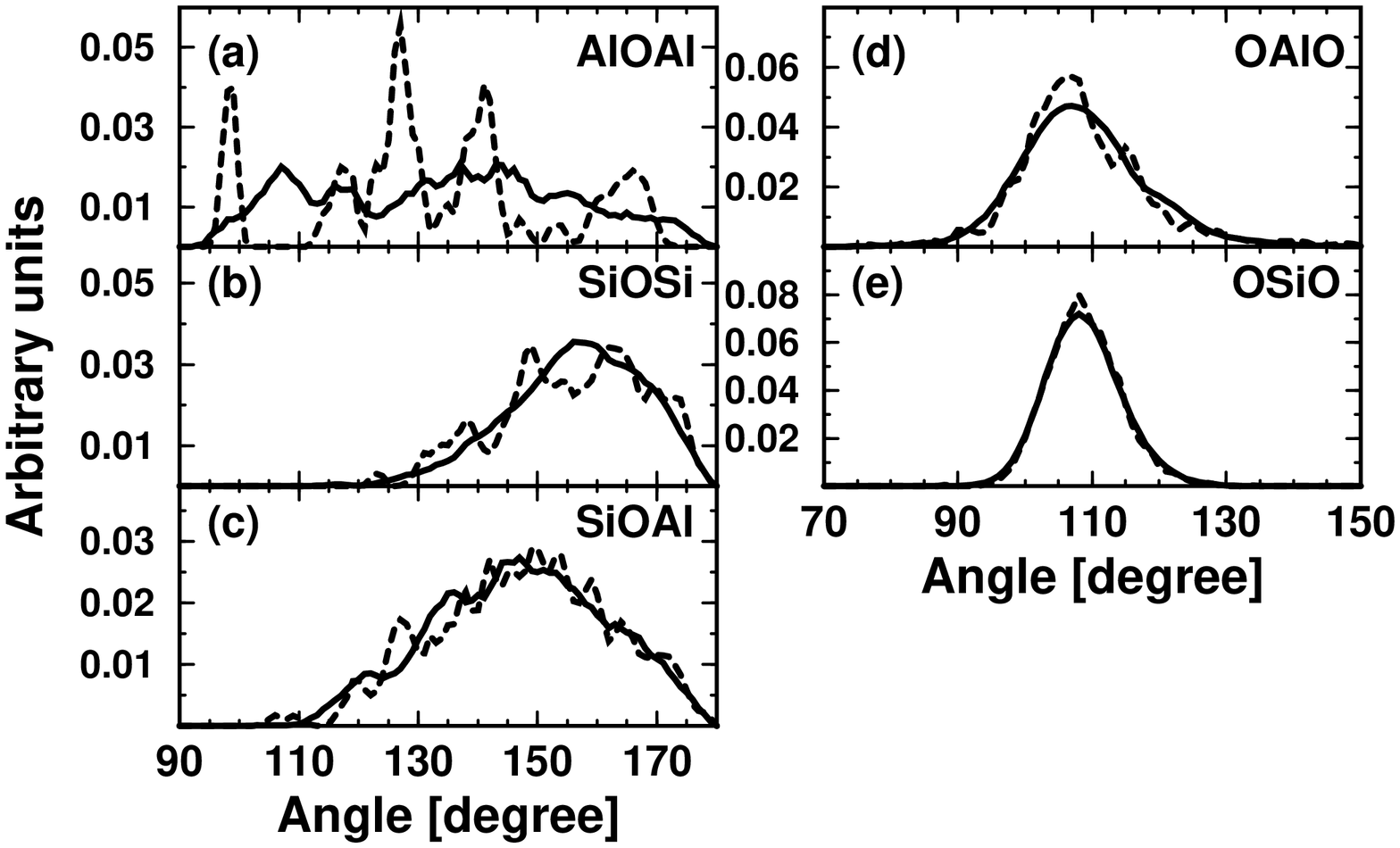}
\caption{ }
\label{figure6}
\end{center}
\end{figure}

\newpage

\begin{figure}
\begin{center}
\includegraphics[width=18cm]{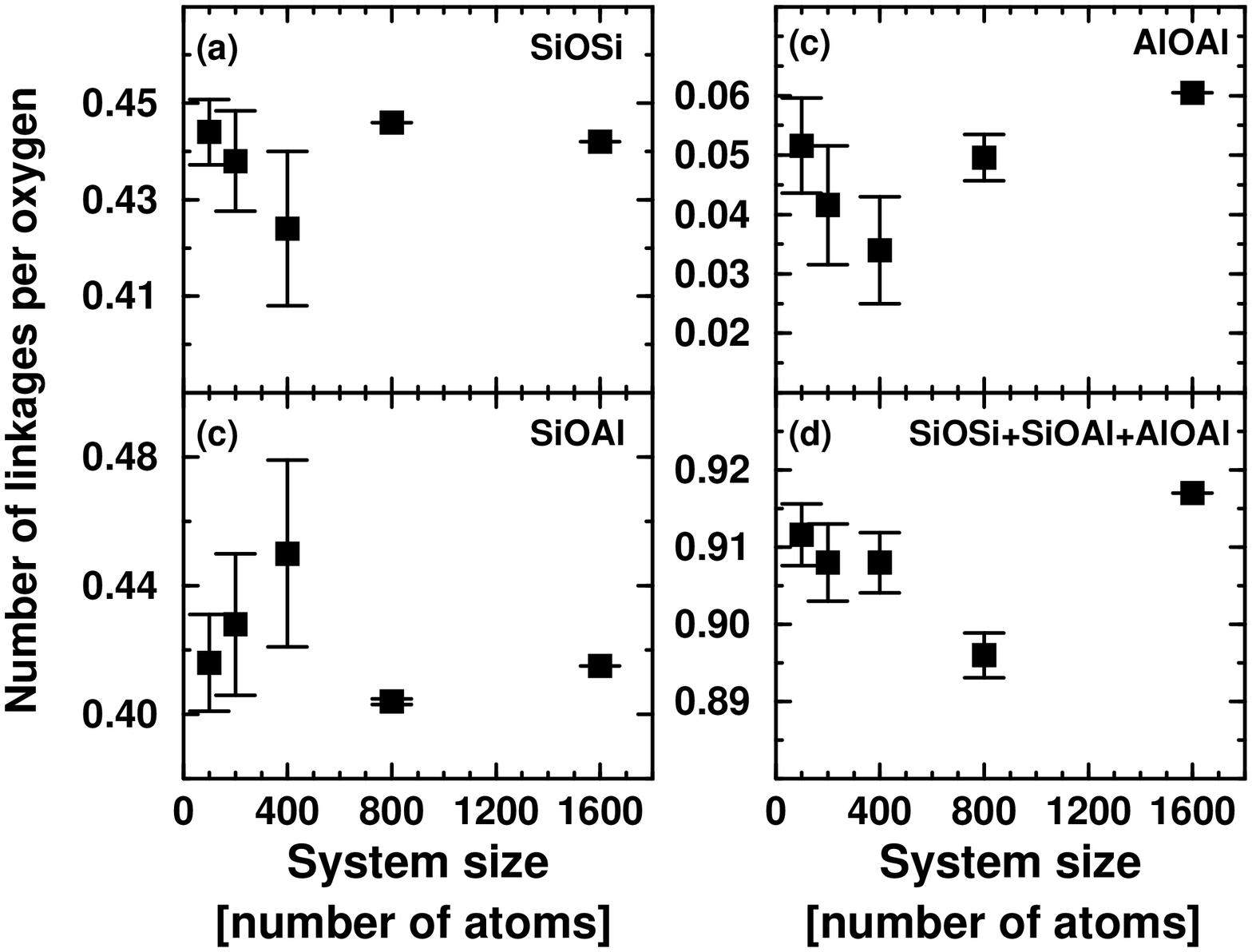}
\caption{ }
 \label{figure7}
\end{center}
\end{figure}

\newpage

\begin{figure}
\begin{center}
\includegraphics[width=18cm]{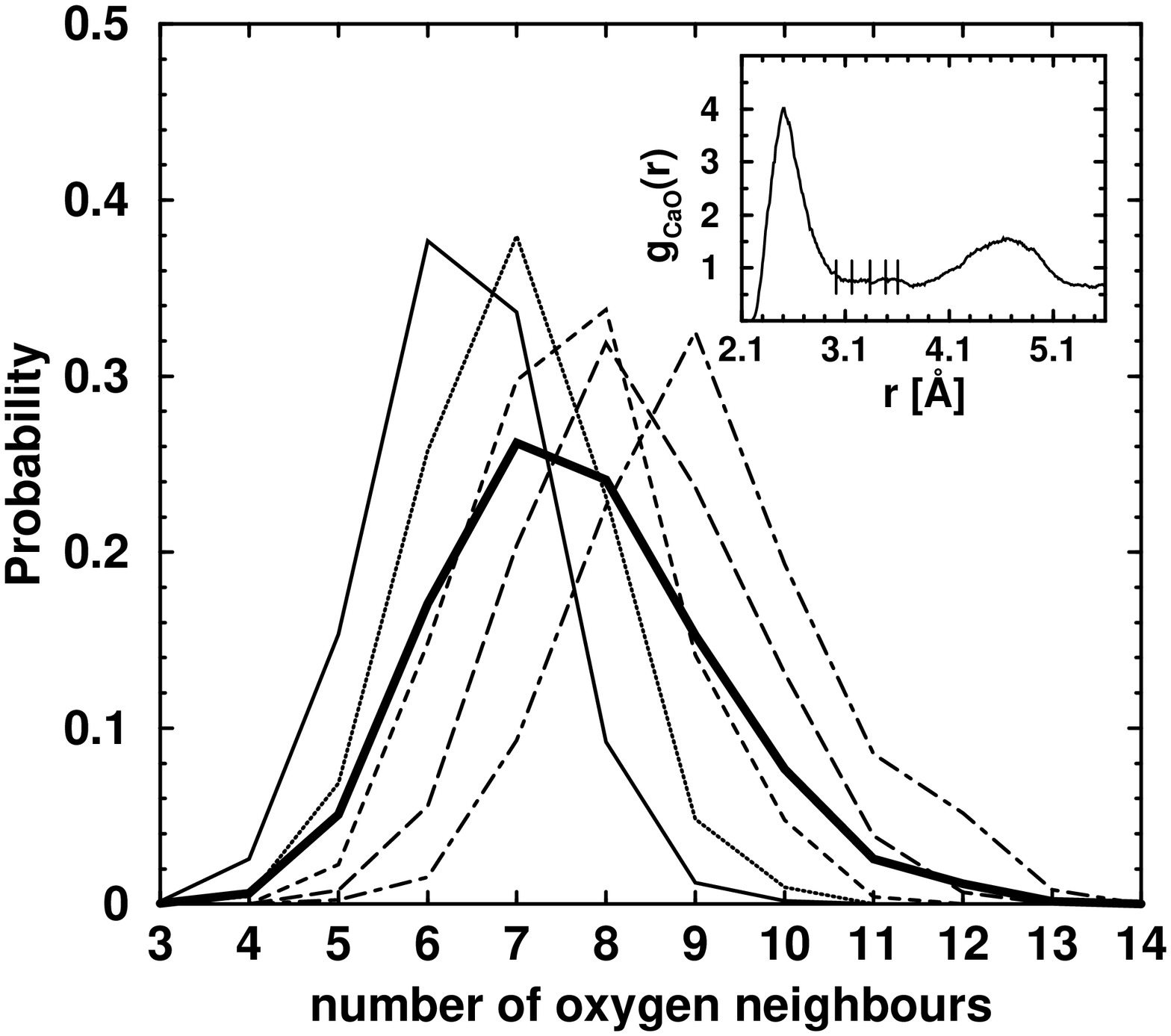}
\caption{ } 
\label{figure8}
\end{center}
\end{figure}

\newpage

\begin{figure}
\begin{center}
\includegraphics[width=18cm]{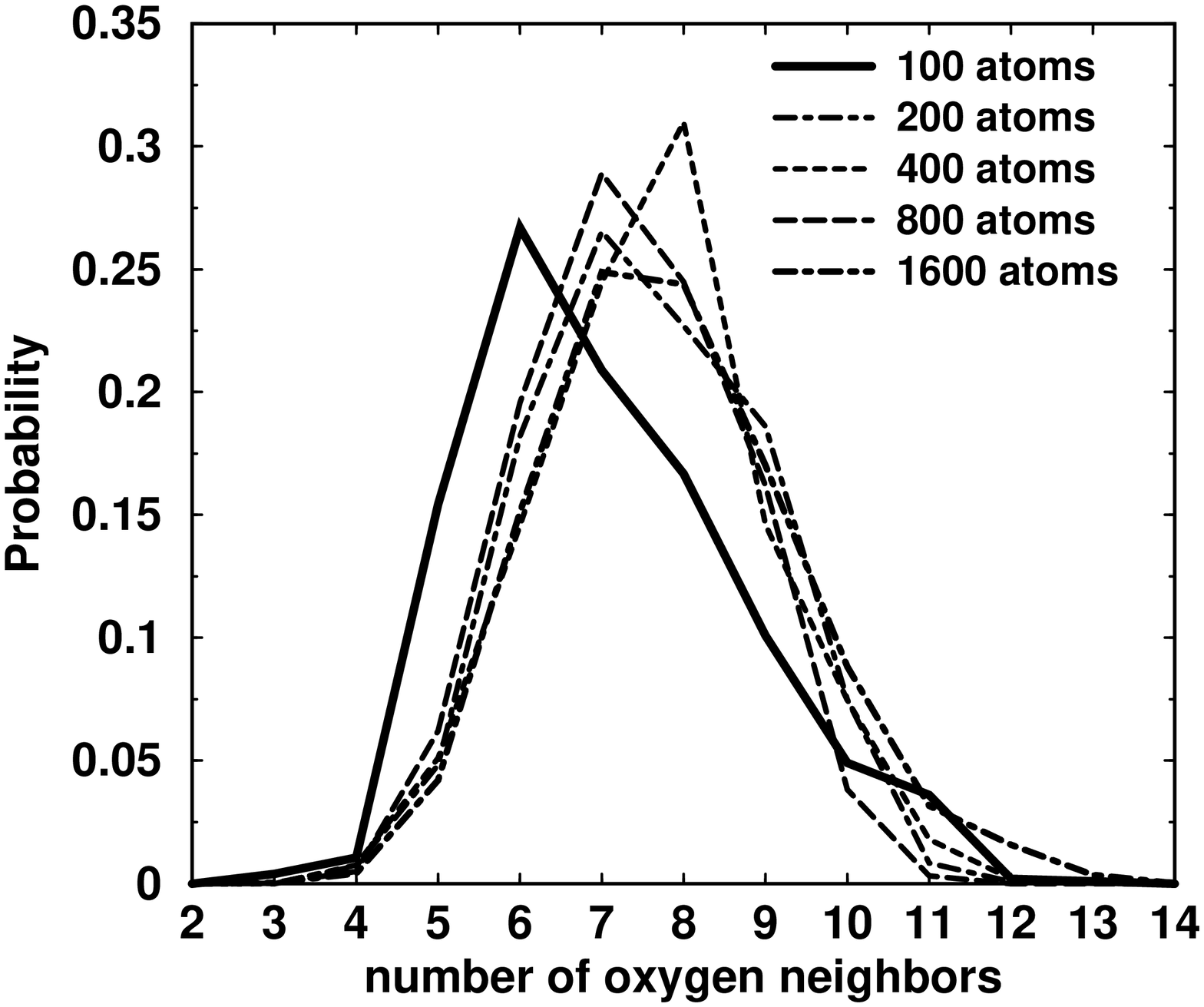}
\caption{ }
 \label{figure9}
\end{center}
\end{figure}

\newpage

\begin{figure}
\begin{center}
\includegraphics[width=18cm]{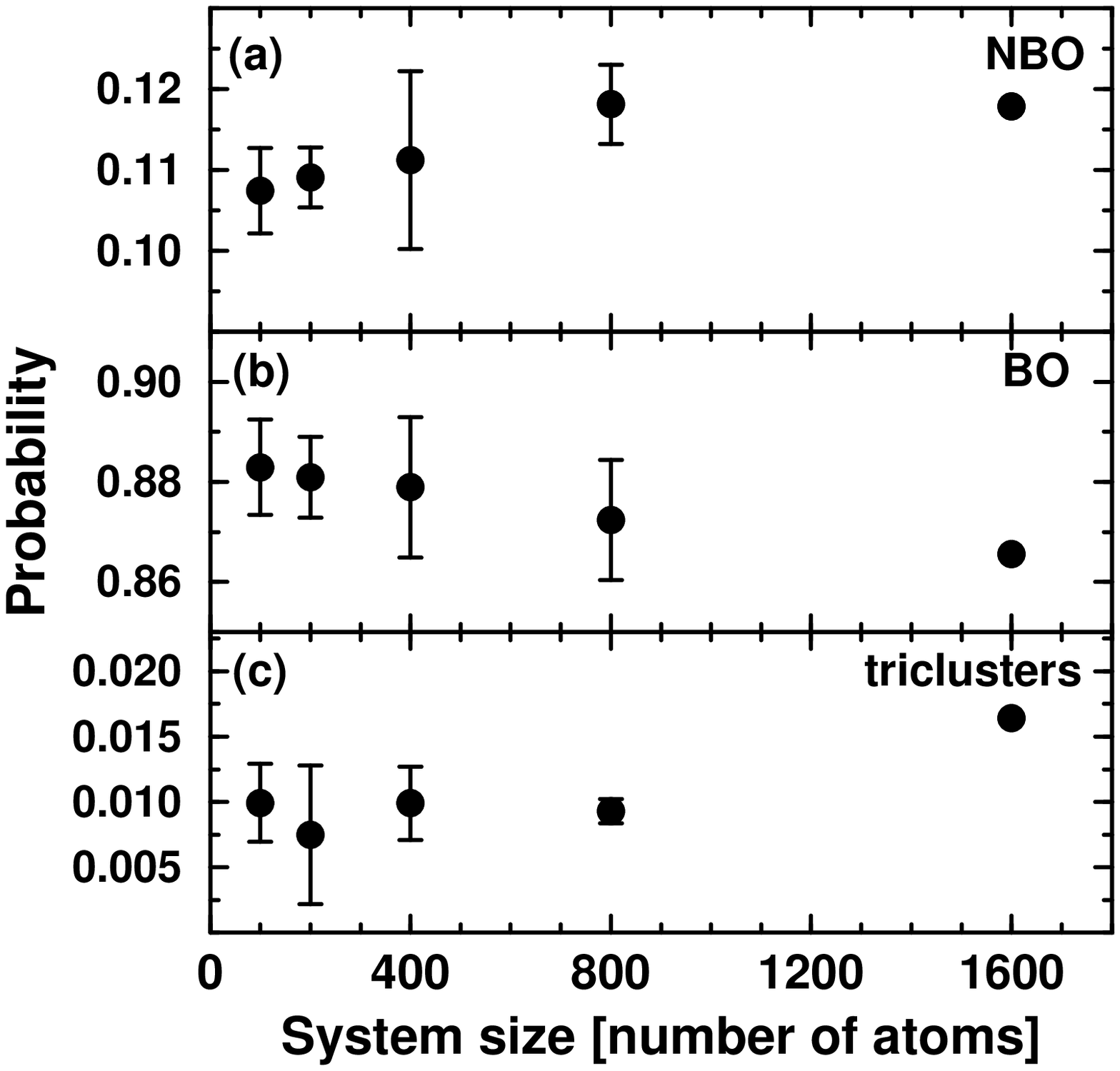}
\caption{ }
 \label{figure10}
\end{center}
\end{figure}

\newpage

\begin{figure}
\begin{center}
\includegraphics[width=18cm]{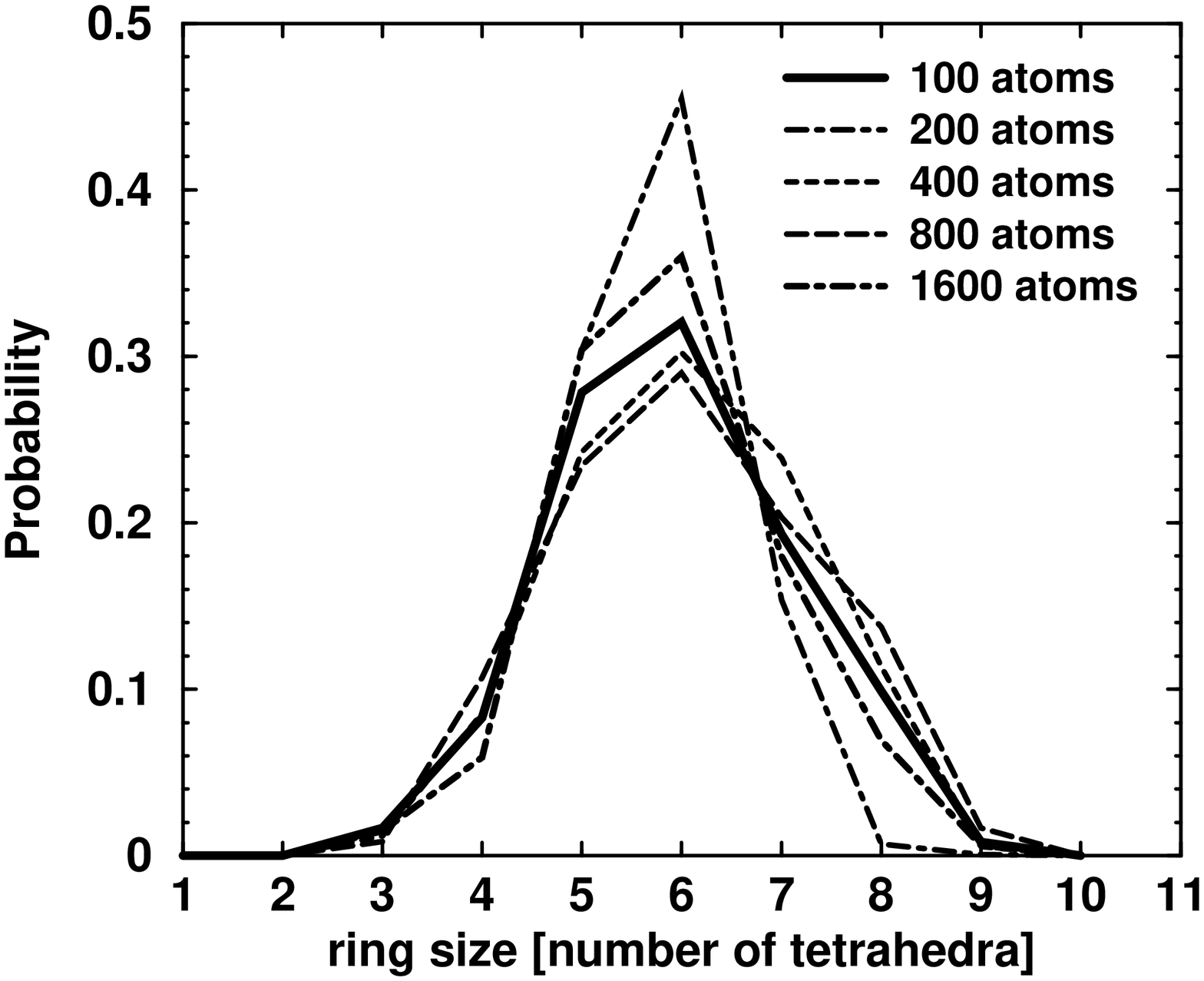}
\caption{ }
\label{figure11}
\end{center}
\end{figure}

\newpage

\begin{figure}
\begin{center}
\includegraphics[width=18cm]{figure12.eps}
\caption{ }
\label{figure12}
\end{center}
\end{figure}

\newpage
\begin{figure}
\begin{center}
\includegraphics[width=18cm]{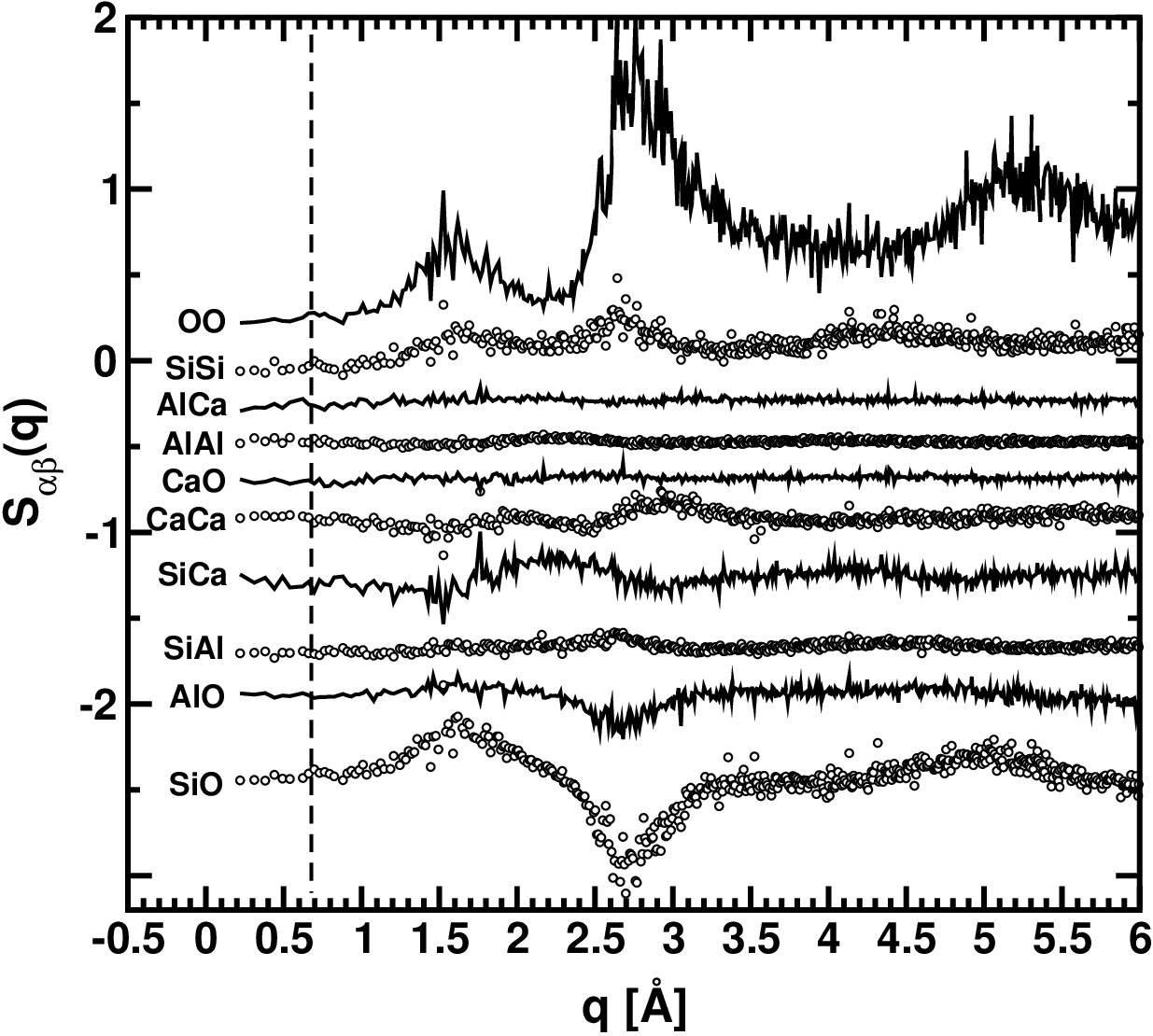}
\caption{ } 
\label{figure13}
\end{center}
\end{figure}

\begin{figure}
\begin{center}
\includegraphics[width=18cm]{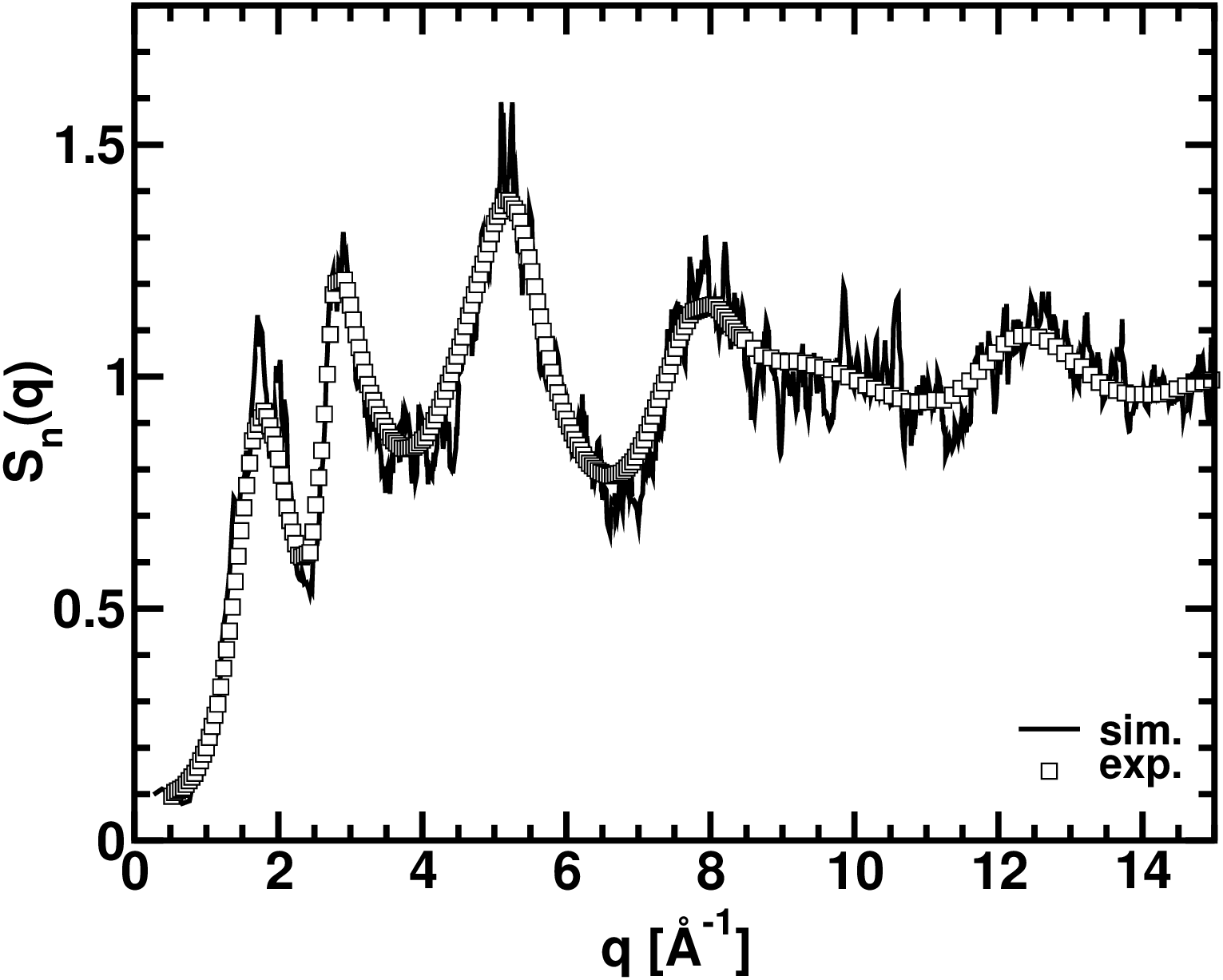}
\caption{ } 
\label{figure14}
\end{center}
\end{figure}


\begin{thebibliography}{99}


\bibitem{aagard}  P. Aagaard and H.C. Helgeson, Am. J. Science {\bf 282}, 237 (1982).
\bibitem{Oelkers} E.H. Oelkers, Geochim. and Cosmochim. Acta  {\bf 65}, 3703 (2001).
\bibitem{grambow} B. Grambow, Mat. Res. Soc. Symp. Proc. {\bf 44}, 15 (1984). 
\bibitem{boksay1} Z. Boksay, G. Bouquet, and S. Dobos, Physics and Chemistry of Glasses {\bf 8}, 140 (1967). 
\bibitem{boksay2} Z. Boksay, G. Bouquet, and S. Dobos, Physics and Chemistry of Glasses {\bf 9}, 69 (1968). 
\bibitem{Kohn}   S.C. Kohn, Mineral. Mag. {\bf 64}, 389 (2000). 
\bibitem{pacaud}  F. Pacaud, N. Jacquet-Francillon, A. Terki and C. Fille, Mat. Res. Soc. Symp. Proc. {\bf 127},  105 (1988).  
\bibitem{vernaz1} E.Y. Vernaz and J.L. Dussossoy, Appl. Geochem. Suppl. {\bf 1}, 13 (1992). 
\bibitem{vernaz2} E.Y. Vernaz and N. Godon, Mat. Res. Soc. Symp. Proc. {\bf 257}, 37 (1992).  
\bibitem{xing} S.-B. Xing, A.C. Buechele, and I.L. Pegg, Mat. Res. Soc. Symp. Proc. {\bf 333}, 541 (1994).


\bibitem{dran} J.C. Dran, J.C. Petit, and C. Brousse, Nature {\bf 319}, 485 (1986).
\bibitem{gin1} S. Gin, Mat. Res. Soc. Symp. Proc. {\bf 663}, 207 (2000).
\bibitem{angeli1} F. Angeli, D. Boscarino, S. Gin, G. Della Mea, B. Boizot, and J.-C. Petit, Physics and Chemistry of Glasses {\bf 42},  279 (2001).
\bibitem{tossel} J.A. Tossel and G. Saghi-Szabo, Geochim. Cosmochim. Acta  {\bf 61}, 1171 (1997).

\bibitem{wu}    Z. Wu {\it et al.}, Phys. Rev. B {\bf 60}, 9216 (1999).
\bibitem{petkov2}   V. Petkov, S.J.L. Billinge, S.D. Shastri, B. Himmel, Phys. Rev. Lett. {\bf 85}, 3436 (2000).




\bibitem{cormier}  L. Cormier, D.R. Neuville, and G. Calas, J. Non-Cryst. Solids {\bf 274}, 110 (2000).

\bibitem{steb1} J.F. Stebbins and Z. Xu, Nature {\bf 390}, 60 (1997).
\bibitem{meyers}   E.R. Meyers, V. Heine, and M.T. Dove, Physics and Chemistry of Minerals {\bf 25}, 457 (1998).
\bibitem{steb2}    J.F. Stebbins, Nature {\bf 330}, 13 (1987).
\bibitem{sio2_epjb} M. Benoit, S. Ispas, P. Jund, and R. Jullien, Euro. Phys. J. B {\bf 13}, 631 (2000). 
\bibitem{ns4_jncs} S. Ispas, M. Benoit, P. Jund and R. Jullien, Phys. Rev. B {\bf 64}, 214206 (2001).

\bibitem{cheeseman}  P. Cheeseman, L.V. Woodcock, and C.A. Angell, J. Chem. Phys. {\bf 65}, 1565 (1976).
\bibitem{soule}      T.F. Soules, J. of Non-Cryst. Solids {\bf 49}, 29 (1982).
\bibitem{stillinger} F.H. Stillinger and T.A. Weber, Phys. Rev. A {\bf 31}, 4234 (1985).
\bibitem{delaye} J.M. Delaye, L. Cormier, D. Ghaleb, G. Calas, J. Non-Cryst. Solids {\bf 290}, 293 (2001).
\bibitem{Doremus} N.P. Bansal and R.H. Doremus, {\it Handbook of Glass Properties}, Academic Press. Inc.
\bibitem{Huang_jncs91} C. Huang and E.C. Behrman, J. Non-Cryst. Solids \textbf{128}, 310 (1991).

\bibitem{winkler} A. Winkler, J. Horbach, W. Kob, K. Binder, J. Chem. Phys. {\bf 120}, 384 (2004).


\bibitem{heuer} B. Doliwa and A. Heuer, J. of Phys.: Cond. Mat. {\bf 15}, S849 (2003).
\bibitem{kim} K. Kim and R. Yamamoto, Phys. Rev. E {\bf 61}, R41 (2000).
\bibitem{horbach96} J. Horbach, W. Kob, K. Binder, and C.A. Angell, Phys. Rev. E {\bf 54}, R5897 (1996).
\bibitem{Hansen} J.-P. Hansen and I.R. McDonald, {\it Theory of Simple Liquids} (Academic, London, 1986).

\bibitem{petkov1} V. Petkov, Th. Gerber, and B. Himmel, Phys. Rev. B {\bf 58}, 11982 (1998).

\bibitem{ganster_thesis} P. Ganster, Ph. D. thesis, Universit\'e Montpellier II (2004).

\bibitem{himmel} B. Himmel, J. Weigelt, Th. Gerber, and M. Nofz, J. Non-Cryst. Solids {\bf 136}, 27 (1991). 

\bibitem{mcmillan} P. McMillan, B. Piriou, and A. Navrotsky, Geochim. Cosmochim. acta {\bf 46}, 2021 (1982). 

\bibitem{calas} G. Calas, G.E. Brown, G.A. Waychunas, and J. Petiau, Phys. and Chem. of Minerals {\bf 15}, 19 (1987).

\bibitem{pettifer} R.F. Pettifer, R. Dupree, I. Farman, and U. Sternberg, J. Non-Cryst. Solids {\bf 106}, 408 (1988).

\bibitem{xiao} Y. Xiao and A.L. Lasaga, Geochim. Cosmochim. Acta {\bf 58}(24), 5379 (1994).

\bibitem{benoit}  M. Benoit, S. Ispas, and M. Tuckerman, Phys. Rev. B {\bf 64}, 224205 (2001).

\bibitem{vollmayr}  K. Vollmayr, W. Kob, and K. Binder, Phys. Rev. B {\bf 54}, 15808 (1996). 

\bibitem{Ni95} www.ncnr.nist.gov/resources/n-lenths/elements/


\bibitem{NS_horbach} J. Horbach, W. Kob, and K. Binder, J. Chem. Geol. {\bf 174}, 87 (2001).

\bibitem{steblee} J.F. Stebbins, S.K. Lee, and J.V. Oglesby, Am. Mineral. {\bf 84}, 983 (1999).

\bibitem{lacy} E.D. Lacy, Physics and Chemistry of Glasses {\bf 4}, 234 (1963).

\bibitem{stebga} J.F. Stebbins and Z. Xu, Science {\bf 390}, 60 (1997).

\bibitem{nevinspera} D. Nevins and F.J. Spera, Am. Mineral. {\bf 83}, 1220 (1998).

\bibitem{BKS} B.W.H. van Beest, G.J. Kramer, and R.A. van Santen, Phys. Rev. Lett. {\bf 64}, 1955 (1990).


\end{thebibliography}
\end{document}